\documentclass[onecolumn,showpacs,preprintnumbers,amsmath,amssymb,showkeys,12ft]{revtex4}

\usepackage{graphicx}
\usepackage{dcolumn}
\usepackage{bm}

\begin{document}
\draft
\date{\today}
\title{Is Sharma-Mittal entropy really a step beyond Tsallis and R\'{e}nyi entropies?}
\author{E. Akt\"{u}rk} \affiliation{Department of Physics Engineering, University of Hacettepe, 06800, Ankara,Turkey}
\author{G. B. Ba\u{g}c\i} \thanks{Corresponding Author}
\email{gbb0002@unt.edu}
\address {Department of Physics, University of North Texas, P.O. Box 311427, Denton, TX 76203-1427,
USA}
\author{R. Sever} \affiliation{Department of Physics, Middle East Technical University, 06800, Ankara,Turkey}
\date{\today}
\pagenumbering{arabic}

\begin{abstract}
We studied the Sharma-Mittal relative entropy and showed that its
physical meaning is the free energy difference between the
off-equilibrium and equilibrium distributions. Unfortunately,
Sharma-Mittal relative entropy may acquire this physical
interpretation only in the limiting case when both parameters
approach to 1 in which case it approaches Kullback-Leibler entropy.
We also note that this is exactly how R\'{e}nyi relative entropy
behaves in the thermostatistical framework thereby suggesting that
Sharma-Mittal entropy must be thought to be a step beyond not both
Tsallis and R\'{e}nyi entropies but rather only as a generalization
of R\'{e}nyi entropy from a thermostatistical point of view. Lastly,
we note that neither of them conforms to the Shore-Johnson theorem
which is satisfied by Kullback-Leibler entropy and one of the
Tsallis relative entropies.
\end{abstract}

  \narrowtext
\newpage \setcounter{page}{1}

\maketitle

\section{\protect\bigskip Introduction}

\bigskip Recently there has been a growing interest in generalized
entropies such as Tsallis [1] and R\'{e}nyi [2] entropies in the
context of a generalized thermostatistics. For example, Tsallis
entropy has been applied to nonlinear diffusion equations [3] and
Fokker-Planck systems [4, 5] while R\'{e}nyi entropy has been
studied for its inverse power law equilibrium distribution [6] and
has been shown to satisfy the zeroth law of thermodynamics [7]. It
has been observed that an unifying entropy exists which seems to
generalize both of these entropies [8, 9, 10]. This two-parametric
entropy is called Sharma-Mittal entropy [11]. It generalizes
Tsallis, R\'{e}nyi and Boltzmann-Gibbs (BG) entropies since its two
parameters generate these entropies in limiting cases. Our aim in
this paper is to shed some light on this new entropy measure from a
thermostatistical point of view. It is organized as follows: In
Section II, the review of the physical meaning of BG relative
entropy i.e., Kullback-Leibler entropy will be given. It will be
seen that it provides a measure of free energy differences between
the off-equilibrium and equilibrium distributions. In Section III,
we will briefly mention Sharma-Mittal entropy and some of its
important properties. In Section IV, we will study the relative
entropy associated with this entropy and show that it can be given a
physical meaning in a generalized thermostatistical framework in
terms of free energy differences only when Sharma-Mittal relative
entropy reduces to Kullback-Leibler entropy. Its connection with
R\'{e}nyi relative entropy will be investigated and shown to behave
in a similar manner. Lastly, in Section V, we will see that
Sharma-Mittal entropy does not conform to Shore-Johnson theorem
sharing also this feature with R\'{e}nyi relative entropy. We will
summarize the conclusions in Section VI.

\bigskip

\bigskip

\section{The Physical meaning of kullback-leibler entropy}

The relative entropy is an important concept whose uses range from
the numerical analysis of protein sequences [12], pricing models in
the market [13], to medical decision making [14]. In this paper, we
will study it from a thermostatistical point of view. In order to
study the physical meaning of any relative entropy in a
thermostatistical framework, one has first to obtain the equilibrium
distribution associated with the entropy of that particular
thermostatistics. This can be done by maximizing BG entropy subject
to some constraints following the well known recipe of entropy
maximization. BG entropy reads

\begin{equation}
S_{BG}(p)=-\sum\limits_{i}^{W}p_{i}\ln p_{i},
\end{equation}

where p$_{\text{i }}$ is the probability of the system in the ith
microstate, W is the total number of the configurations of the
system. Note that Boltzmann constant k is taken to be equal to one
throughout the paper. Let us assume that the internal energy
function is given by $U=\sum\limits_{i}\varepsilon _{i}p_{i}$ where
$\varepsilon_{i}$ denotes the energy of the ith microstate. In order
to get the equilibrium distribution associated with BG entropy, we
maximize the following functional

\begin{equation}
\Phi (p)=-\sum\limits_{i}^{W}p_{i}\ln p_{i}-\alpha
\sum\limits_{i}^{W}p_{i}-\beta
\sum\limits_{i}^{W}\varepsilon_{i}p_{i},
\end{equation}

where $\alpha$ and $\beta$ are Lagrange multipliers related to
normalization and internal energy constraints respectively. Equating
the derivative of the functional to zero, we obtain
\begin{equation}
\frac{\delta \Phi (p)}{\delta p_{i}}=-\ln \widetilde{p}_{i}-1-\alpha
-\beta \varepsilon_{i}=0.
\end{equation}

Tilde denotes the equilibrium distribution obtained by the
maximization of BG entropy. By multiplying Eq. (3) by
$\widetilde{p}_{i}$ and summing over i, using the normalization and
internal energy constraints, we have

\begin{equation}
\alpha +1=\widetilde{S}_{BG}-\beta \widetilde{U}.
\end{equation}

Substitution of Eq. (4) into Eq. (3) results in the following
equilibrium distribution

\bigskip
\begin{equation}
\widetilde{p}_{i}=e^{-\widetilde{S}_{BG}}e^{\beta
\widetilde{U}}e^{-\beta \varepsilon_{i}}.
\end{equation}

The relative entropy in ordinary BG case is called Kullback-Leibler
(K-L) entropy [15]. It reads

\begin{equation}
K[p\Vert q]\equiv \sum_{i}p_{i}\ln (p_{i}/q_{i}).
\end{equation}

Note that it is a convex function of ${p}_{i}$, always non-negative
and equal to zero if and only if $p=q$. If we now use the
equilibrium distribution $\widetilde{p}$ as the reference
distribution in K-L entropy, we can write

\begin{equation}
K[p\Vert \widetilde{p}]=\sum_{i}p_{i}\ln (p_{i}/\widetilde{p}_{i}).
\end{equation}

The equation above can be rewritten as

\begin{equation}
K[p\Vert \widetilde{p}]= -S_{BG}-\sum_{i}p_{i}\ln \widetilde{p}_{i}.
\end{equation}

We then insert the equilibrium distribution given by Eq. (5) in the
equation above to find

\begin{equation}
K[p\Vert \widetilde{p}]=
-S_{BG}-\sum_{i}p_{i}(-\widetilde{S}_{BG}+\beta \widetilde{U}-\beta
\varepsilon_{i}).
\end{equation}

Taking care of the effect of summation, we have

\begin{equation}
K[p\Vert \widetilde{p}]= -S_{BG}+\widetilde{S}_{BG}-\beta
\widetilde{U}+\beta U,
\end{equation}

which can be cast into the form

\begin{equation}
K[p\Vert \widetilde{p}]= \beta (F-\widetilde{F}).
\end{equation}

The free energy term is given as usual by $F=U-S_{BG}/\beta$. The
result above shows us that the physical meaning of the K-L entropy
is nothing but the difference of the off-equilibrium and equilibrium
free energies when the reference distribution is taken to be the
equilibrium distribution given by Eq. (5) above.  This result can be
used, for example, to study equilibrium fluctuations or
non-equilibrium relaxation of polymer chains [16].

\section{Sharma-Mittal Entropy}

In this Section, we will briefly review the Sharma-Mittal entropy
and some of its important properties from a generalized
thermostatistical point of view. The Sharma-Mittal entropy [11] is
given by

\begin{equation}
S_{SM}(p)=\frac{1}{1-r}\Big[\Big(\sum_{i}p_{i}^{q}\Big)^{(\frac{1-r}{1-q})}-1\Big].
\end{equation}

In the limit r$\rightarrow 1$, Sharma-Mittal entropy becomes
R\'{e}nyi entropy [2] which is

\begin{equation}
S_{R}(p)=\frac{1}{1-q}\ln (\sum\limits_{i}p_{i}^{q}),
\end{equation}

while for r$\rightarrow q$, it is Tsallis entropy [1] given by

\begin{equation}
S_{T}(p)=\frac{\sum_{i}p_{i}^{q}-1}{1-q}.
\end{equation}

In the limiting case when both parameters approach 1, we recover the
ordinary Boltzmann-Gibbs (BG) entropy which reads

\begin{equation}
S_{BG} (p)=-\sum_{i}p\ln p_{i}.
\end{equation}

For two statistically independent systems given by probability
distributions ${p_{i}}$ and ${p'_{k}}$, Sharma-Mittal entropy
satisfies [9]

\begin{equation}
S_{SM}(\{p_{i}p_{k}^{\prime
}\})=S_{SM}(\{p_{i}\})+S_{SM}(\{p_{k}^{\prime
}\})+(1-r)S_{SM}(\{p_{i}\})S_{SM}(\{p_{k}^{\prime }\}).
\end{equation}

This relation is important in order to see how Sharma-Mittal measure
includes both Tsallis and R\'{e}nyi entropies as its limiting cases.
When we take $r=q$, it becomes the relation satisfied by Tsallis
entropy and shows its nonextensivity. On the other hand, when $r=1$,
we have the extensive property which relates to R\'{e}nyi entropy.
In this sense, Sharma-Mittal entropy includes both extensive and
nonextensive features in it.

One feature of Sharma-Mittal entropy is that it fails to be concave
[8]. The concavity entails thermodynamic stability and can be
defined as follows: Consider probability distributions
$P=\{p_{1},p_{2},...,p_{N}\}$ and $P^{\prime }=\{p_{1}^{\prime
},p_{2}^{\prime },...,p_{N}^{\prime }\}$. Let us also define an
intermediate probability distribution given by $P^{\prime \prime
}=\{p_{1}^{\prime \prime },p_{2}^{\prime \prime },...p_{N}^{\prime
\prime }\}$ where

\begin{equation}
p_{i}^{\prime \prime }\equiv \mu p_{i}+(1-\mu )p_{i}^{\prime
},\forall \mu \in \lbrack 0,1].
\end{equation}

Now, S(P) is said to be concave if and only if

\begin{equation}
S(P^{\prime \prime })\geqslant \mu S(P)+(1-\mu )S(P^{\prime }).
\end{equation}

It is worth remark that BG entropy and Tsallis entropy is concave
whereas R\'{e}nyi entropy is not concave for all values of parameter
$q$ [17].

Another feature of Sharma-Mittal entropy is that it is not
Lesche-stable [18]. Lesche stability checks the stability of the
entropy functional under arbitrary small variations of the
probabilities. It can be stated in a more rigorous way by defining
the deformation of the probability distribution as

\begin{equation}
\left\Vert p-p^{\prime }\right\Vert = \sum_{i}\left\vert
p_{i}-p_{i}^{\prime }\right\vert <\delta _{\varepsilon },\forall
\delta _{\varepsilon }>0.
\end{equation}

Then, S(P) is Lesche-stable if

\begin{equation}
\Delta =\left\vert \frac{S(P)-S(P^{\prime })}{S_{\max }}\right\vert
<\varepsilon ,\forall \varepsilon >0,
\end{equation}

where $S_{max}$ is the maximum value that S can attain over all
microstates. It has been shown that BG and Tsallis entropy is always
Lesche-stable whereas R\'{e}nyi entropy is unstable for all $q\neq1$
[17, 18].

Lastly, it may be remarked that Sharma-Mittal entropy does not lead
to finite entropy production per unit time whereas BG and Tsallis
entropies do [8].

\section{Sharma-Mittal Relative Entropy AND FREE ENERGY}

In order to study the physical meaning of Sharma-Mittal relative
entropy in thermostatistical framework as we did with BG entropy in
Section II, we have to obtain the equilibrium distribution
corresponding to Sharma-Mittal entropy. For this purpose, we begin
by maximizing the following associated functional

\begin{equation}
\Phi_{SM}(p)=\frac{1}{1-r}\Big[\Big(\sum_{i}^{W}p_{i}^{q}\Big)^{(\frac{1-r}{1-q})}-1\Big]-\alpha\sum_{i}^{W}p_{i}-\beta
\frac{\sum_{i}^{W}\varepsilon_{i}p_{i}^{q}}{\sum_{i}^{W}p_{i}^{q}}.
\end{equation}

We take the derivative of the functional and equal it to zero in
order to obtain the following

\begin{equation}
\frac{\delta\Phi_{SM}(p)}{\delta p_i}=
\frac{q}{1-q}\tilde{p_{i}}^{q-1}\Big(\sum_{i}^{W}\tilde{p_{i}}^{q}\Big)^{(\frac{q-r}{1-q})}-\alpha-{\beta}^{*}q
\tilde{p_{i}}^{q-1}(\varepsilon_i-\tilde{U})=0,
\end{equation}

where $\beta^{*}$ is given by

\begin{equation}
\beta^{*}=\frac{\beta}{\Big(\sum_{i} \tilde{p_i}^q\Big)}.
\end{equation}

Multiplying Eq. (22) by $\tilde{p_i}$ and summing over the index i,
we obtain

\begin{equation}
\alpha=\frac{q}{1-q}\Big(\sum_{i}^{W}\tilde{p_{i}}^{q}\Big)^{(\frac{1-r}{1-q})}.
\end{equation}

Note that tilde shows that the distribution is calculated at
equilibrium. Substituting this explicit expression of $\alpha$ into
Eq.(22), we calculate the associated equilibrium distribution [9] as

\begin{equation}
\tilde{p_i}=\Big(\frac{1}{\sum\tilde{p}_i^q}\Big)^{\frac{1}{1-q}}\Big[1-(1-q)\beta^{**}(\varepsilon_i-\tilde{U})\Big]^{\frac{1}{1-q}},
\end{equation}

where $\beta^{**}$ is given by

\begin{equation}
\beta^{**}=\frac{\beta^{*}}{(\sum_{i}\tilde{p}_{i}^q)^{\frac{q-r}{1-q}}}=\frac{\beta}{(\sum_{i}\tilde{p}_i^q)^{\frac{1-r}{1-q}}}.
\end{equation}

From Eq.(12), we see that

\begin{equation}
\Big(\sum_{i}p_{i}^{q}\Big)^{(\frac{1-r}{1-q})}=1+(1-r)S_{SM},
\end{equation}

which also holds for equilibrium distribution given by Eq.(25)

\begin{equation}
\Big(\sum_{i}\tilde{p_{i}}^{q}\Big)^{(\frac{1-r}{1-q})}=1+(1-r)\tilde{S}_{SM}.
\end{equation}

The Sharma-Mittal divergence is given by [19]

\begin{equation}
I_{SM}[p\parallel\tilde{p}]=\frac{1}{r-1}\Big[\Big(\sum_{i}p_{i}^{q}\tilde{p_{i}}^{1-q}\Big)^{(\frac{1-r}{1-q})}-1\Big].
\end{equation}

We now substitute equilibrium distribution in Eq. (25) as the
reference distribution into Sharma-Mittal divergence given above and
obtain

\begin{equation}
I_{SM}[p\parallel\tilde{p}]=\frac{1}{r-1}\Big[\Big\{\sum_{i}p_{i}^{q}(1+(1-r)\tilde{S}_{SM})^{\frac{q-1}{1-r}}
[1-(1-q)\beta^{**}(\varepsilon_i-\tilde{U})]\Big\}^{(\frac{1-r}{1-q})}-1\Big].
\end{equation}

Using Eq.(27), it can be put into a more appropriate form which is

\begin{equation}
I_{SM}[p\Vert \widetilde{p}]=\frac{1}{r-1}[\{\frac{1+(1-r)S_{SM}}{1+(1-r)%
\widetilde{S}_{SM}}\{1-(1-q)\beta ^{\ast \ast }(U-\widetilde{U})\}^{(\frac{%
1-r}{1-q})}\}-1].
\end{equation}

The expression above is very different than Eq. (11). It cannot be
written in terms of free energy differences. Indeed, this can be
achieved only by taking the limit $q$ approaches 1 first

\begin{equation}
I_{SM}[p\Vert \widetilde{p}]=\frac{1}{r-1}[\{\frac{1+(1-r)S_{SM}}{1+(1-r)%
\widetilde{S}_{SM}}e^{\beta ^{\ast \ast
}(\widetilde{U}-U)((1-r)}\}-1],
\end{equation}

where the internal energy functions and $\beta ^{\ast \ast }$ must
be calculated at $q=1$ and then considering the limit of the above
expression as $r$ goes to 1, which in turn gives

\begin{equation}
K[p\Vert \widetilde{p}]= \beta (F-\widetilde{F}).
\end{equation}

 This shows that the physical meaning of Sharma-Mittal divergence
 is the difference between the off-equilibrium free energy and
 equilibrium free energy when the reference distribution is taken
 to be the equilibrium distribution obtained from the maximization
 of Sharma-Mittal entropy only in the limiting case
when both parameters approach to 1 in which case it approaches
Kullback-Leibler entropy.

This negative result above is the one exactly mimicked by the
R\'{e}nyi relative entropy [20]. It reads

\begin{equation}
I_{q}[p\Vert r]=\frac{1}{q-1}\ln
(\sum\limits_{i}p_{i}^{q}r_{i}^{1-q}).
\end{equation}

Note that this definition of R\'{e}nyi relative entropy is always
non-negative and equal to zero if and only if $p=r$. It also reduces
to K-L entropy as the parameter $q$ approaches 1. Let us write the
associated functional where internal energy constraint is given in
terms of escort probabilities i.e.,
$U=\frac{\sum\limits_{i}\varepsilon
_{i}p_{i}^{q}}{\sum\limits_{j}p_{j}^{q}}$, thereby yielding

\begin{equation}
\Phi _{R}(p)=\frac{1}{1-q}\ln (\sum\limits_{i}^{W}p_{i}^{q})-\alpha
\sum\limits_{i}^{W}p_{i}-\beta \frac{\sum\limits_{i}^{W}\varepsilon_{i}p_{i}^{q}}{%
\sum\limits_{i}^{W}p_{j}^{q}}.
\end{equation}

We again take the derivative of the functional and equate it to zero
in order to obtain the following

\begin{equation}
\frac{\delta \Phi _{R}(p)}{\delta p_{i}}=\frac{q}{1-q}\frac{\widetilde{p}%
_{i}^{q-1}}{\sum\limits_{j}\widetilde{p}_{j}^{q}}-\alpha -\beta ^{\ast }q%
\widetilde{p}^{q-1}(\varepsilon_{i}-\widetilde{U})=0,
\end{equation}

where $\beta^{\ast}$ is given by

\begin{equation}
\beta ^{\ast }=\frac{\beta }{\sum\limits_{j}\widetilde{p}_{j}^{q}}.
\end{equation}

Multiplying Eq. (36) by $\widetilde{p}_{i}$ and summing over the
index i, we find

\begin{equation}
\alpha =\frac{q}{1-q}.
\end{equation}

Note that tilde again denotes that the distribution is calculated at
equilibrium. Substituting this explicit expression of $\alpha$ back
into Eq. (36), R\'{e}nyi equilibrium distribution reads

\begin{equation}
\widetilde{p}_{i}=(\frac{1}{e^{(1-q)\widetilde{S}_{R}}}-(1-q)\beta
^{\ast }(\varepsilon _{i}-\widetilde{U}))^{1/(1-q)}.
\end{equation}

Following the same steps as before, we then substitute equilibrium
distribution above as the reference distribution into the R\'{e}nyi
relative entropy given by Eq. (34) and get

\begin{equation}
I_{q}[p\Vert \widetilde{p}]=\frac{1}{q-1}\ln (\sum\limits_{i}p_{i}^{q}((%
\frac{1}{e^{(1-q)\widetilde{S}_{R}}}-(1-q)\beta ^{\ast }(\varepsilon _{i}-%
\widetilde{U})))).
\end{equation}

It can be put into a more appropriate form which is

\begin{equation}
I_{q}[p\Vert \widetilde{p}]=\frac{1}{q-1}\ln (e^{(1-q)(S_{R}-\widetilde{S}%
_{R})}-(1-q)\beta ^{\ast \ast }(U-\widetilde{U})),
\end{equation}

where $\beta ^{\ast \ast }$ is given by

\begin{equation}
\beta ^{\ast \ast }=\frac{\beta }{\sum\limits_{j}\widetilde{p}_{j}^{q}}%
\sum\limits_{i}p_{i}^{q}.
\end{equation}

Making Taylor series expansion about $q=1$ for the exponential term
within the parentheses and keeping the first two terms only, we have

\begin{equation}
I_{q}[p\Vert \widetilde{p}]=\frac{1}{q-1}\ln (1+(1-q)(S_{BG}-\widetilde{S}%
_{BG})-(1-q)\beta ^{\ast \ast }(U-\widetilde{U})).
\end{equation}

Arranging the terms as follows

\begin{equation}
I_{q}[p\Vert \widetilde{p}]=\frac{1}{q-1}\ln [1+(1-q)\{(S_{BG}-\widetilde{S}%
_{BG})-\beta ^{\ast \ast }(U-\widetilde{U})\}]
\end{equation}

and making Taylor expansion about $q=1$ again but this time to the
logarithmic term, we obtain

\begin{equation}
I_{q}[p\Vert \widetilde{p}]=\frac{1}{(q-1)}(1-q)[(S_{BG}-\widetilde{S}%
_{BG})-\beta (U-\widetilde{U})],
\end{equation}

which can be written as

\begin{equation}
I_{q}[p\Vert \widetilde{p}]=\beta \lbrack (U-S_{BG}/\beta )-(\widetilde{U}-%
\widetilde{S}_{BG}/\beta )].
\end{equation}

This is nothing but the free energy differences since it can be
rewritten as

\begin{equation}
I_{q}[p\Vert \widetilde{p}]=\beta(F-\widetilde{F}),
\end{equation}

where free energy expressions are given by exactly as in the BG
case. This is exactly the same expression obtained in Section II by
using BG entropy and K-L entropy. It should be noted that the first
Taylor expansion turned the R\'{e}nyi entropies into BG entropies
while second Taylor expansion turned the Lagrange multiplier and
internal energy functions into their corresponding BG values [21].

This shows that Sharma-Mittal relative entropy behaves exactly in
the same way as R\'{e}nyi relative entropy when one considers them
in terms of their physical meanings in a generalized
thermostatistical framework.

\section{\protect\bigskip Sharma-Mittal Relative Entropy and Shore-Johnson Theorem}

At this point, it is important to remember Shore-Johnson theorem
(see Refs. [22, 23] for details). According to it, any relative
entropy $J[p\Vert r]$ with the prior $r_{i}$ and posterior $p_{i}$
which satisfies five very general axioms, must be of the form

\begin{equation}
J[p\Vert r]=\sum\limits_{i}p_{i}h(p_{i}/r_{i}),
\end{equation}

for some function $h(x)$. These axioms are listed as

\begin{enumerate}
  \item Axiom of Uniqueness: If the same problem is solved twice,
  then the same answer is expected to result both times.
  \item Axiom of Invariance: The same answer is expected when the
  same problem is solved in two different coordinate systems, in
  which the posteriors in the two systems should be related by the
  coordinate transformation.
  \item Axiom of System Independence: It should not matter whether
  one accounts for independent information about independent systems
  separately in terms of their marginal distributions or in terms of
  the joint distribution.
  \item Axiom of Subset Independence: It should not matter whether
  one treats independent subsets of the states of the systems in
  terms of their separate conditional distributions or in terms of
  the joint distribution.
  \item Axiom of Expansibility: In the absence of new information,
  the prior should not be changed.
\end{enumerate}

Ordinary relative entropy i.e., K-L entropy is in accordance with
Shore-Johnson theorem since we can find a function $h(x)$ which
allows us to write K-L entropy as Eq. (48) requires. This function
$h(x)$ is nothing but natural logarithm indeed. In the case of
R\'{e}nyi relative entropy [20] given by Eq. (34), we see that it
cannot be cast into a form which will conform to the Shore-Johnson
theorem for any function $h(x)$. Inspection of Sharma-Mittal
relative entropy shows that it shares also this feature of R\'{e}nyi
relative entropy. In other words, both fails to conform to
Shore-Johnson theorem. On the other hand, the nonextensive
counterpart of relative entropy given by

\begin{equation}
I_{q}^{Tsallis}[p\Vert r]=\frac{1}{1-q}[1-\sum%
\limits_{i}p_{i}^{q}r_{i}^{1-q}],
\end{equation}

is seen to conform to Shore-Johnson theorem when the function $h(x)$
is taken to be the the negative of the $q$-logarithm function
defined by $\ln _{q}(x)=\frac{x^{1-q}-1}{1-q}$ but with argument $x$
replaced by $1/x$ [23]. Therefore, in the case of nonextensive
thermostatistics, one has a relative entropy which conforms to
Shore-Johnson theorem as K-L entropy in BG thermostatistics does.

\section{RESULTS AND DISCUSSIONS}

The Sharma-Mittal entropy seems to generalize both R\'{e}nyi and
Tsallis entropies through the adjustment of its two parameters. We
have investigated whether this interpretation  is plausible from the
thermostatistical point of view using the associated Sharma-Mittal
relative entropy. The relative entropy is an important concept and
has many applications in diverse fields such as quantum information
theory [24], biophysics [12] and finance [13]. Its physical meaning
in ordinary thermostatistics is the difference of free energies
associated with equilibrium and off-equilibrium distributions. In
this paper, we have shown that a similar result can be obtained in
the case of Sharma-Mittal entropy but only when it reduces to K-L
entropy, rendering the use of Sharma-Mittal relative entropy
redundant in this generalized thermostatistical framework. We also
observe that this is exactly how R\'{e}nyi relative entropy behaves
when it is subject to same kind of calculation. Another negative
feature which Sharma-Mittal relative entropy has in common with
R\'{e}nyi entropy is that associated relative entropies violate the
Shore-Johnson theorem which is satisfied by Kullback-Leibler entropy
and one of the Tsallis relative entropies. Considering all these
negative results common to both of them including the failure of
concavity and stability, we believe that Sharma-Mittal entropy must
be thought to be a step beyond not both Tsallis and R\'{e}nyi
entropies but rather as a generalization of R\'{e}nyi entropy from a
thermostatistical point of view although the explicit form of
Sharma-Mittal entropy suggests that it has both of these entropies
in its content when we only consider the limiting values of its
parameters.

\end{document}